\documentclass[12pt]{article}
\usepackage{amssymb}
\usepackage{array}
\usepackage{float}
\usepackage{epsfig}
\usepackage{dsfont}
\usepackage{amssymb,amsmath}
\usepackage{slashed}
\usepackage{multirow}

\newcommand{\bea}{\begin{eqnarray}}
\newcommand{\eea}{\end{eqnarray}}
\newcommand{\beq}{\begin{equation}}
\newcommand{\eeq}{\end{equation}}

\newcommand{\phie}{\sigma}

 \makeatletter
\def\alt{\mathrel{\mathpalette\gl@align<}}
\def\agt{\mathrel{\mathpalette\gl@align>}}
\def\gl@align#1#2{\lower.6ex\vbox{\baselineskip\z@skip\lineskip\z@
\ialign{$\m@th#1\hfil##\hfil$\crcr#2\crcr\sim\crcr}}} \makeatother

\begin{document}

\begin{center}
\baselineskip 20pt 
{\Large\bf Tensor to Scalar Ratio 
in Non-Minimal $\phi^4$ Inflation}

\vspace{1cm}

{\large 
Nobuchika Okada$^{a,}$\footnote{ E-mail: okadan@ua.edu}, 
Mansoor Ur Rehman$^{b,}$\footnote{E-Mail: rehman@udel.edu} 
and Qaisar Shafi$^{b,}$\footnote{ E-mail: shafi@bartol.udel.edu} 
} 
\vspace{.5cm}

{\baselineskip 20pt \it
$^a$Department of Physics and Astronomy, University of Alabama, \\ 
Tuscaloosa, AL 35487, USA \\
\vspace{2mm} 
$^b$Bartol Research Institute, Department of Physics and Astronomy, \\
University of Delaware, Newark, DE 19716, USA \\
}

\vspace{1cm}
\end{center}

\begin{abstract}
We reconsider non-minimal $\lambda\,\phi^4$ chaotic inflation 
which includes the gravitational coupling term 
$\xi\,\mathcal{R}\,\phi^2$, where $\phi$ denotes 
a gauge singlet inflaton field and $\mathcal{R}$ 
is the Ricci scalar. 
For $\xi \gg 1$ we require, following recent discussions, 
that the energy scale $\lambda^{1/4} m_P / \sqrt{\xi}$ 
for inflation should not exceed the effective UV cut-off 
scale $m_P / \xi$, where $m_P$ denotes the reduced Planck scale. 
The predictions for the tensor to scalar ratio $r$ and 
the scalar spectral index $n_s$ are found to lie within  
the WMAP 1-$\sigma$ bounds for 
$10^{-12} \lesssim \lambda \lesssim 10^{-4}$ and 
$10^{-3} \lesssim \xi \lesssim 10^2$. 
In contrast, the corresponding predictions 
of minimal $\lambda\,\phi^4$ chaotic inflation  
lie outside the WMAP 2-$\sigma$ bounds.
We also find that $r \gtrsim 0.002$, provided the scalar spectral 
index $n_s \geq 0.96$. 
In estimating the lower bound on $r$ we take into account 
possible modifications due to quantum corrections of 
the tree level inflationary potential.
\end{abstract}

The idea that the inflaton may be a scalar field having an
additional non-minimal coupling to gravity has received 
a fair amount of attention  \cite{Salopek:1988qh}-\cite{Pallis:2010wt}.
In one of the simplest scenarios of this kind, the Standard Model 
(SM) Higgs doublet $H$ has a relatively strong 
non-minimal gravitational interaction 
$\xi\,\mathcal{R}\,H^\dagger H $, where $\mathcal{R}$ is the Ricci 
scalar and $\xi$ a dimensionless coupling 
whose magnitude is estimated to be of order $10^3$-$10^4$ 
based on measurements by WMAP \cite{Komatsu:2010fb} and 
other CMB anisotropy experiments. 
This SM Higgs based inflationary scenario is currently mired 
in some controversy stemming from arguments first put forward in 
\cite{cutoff} that for $\xi \gg 1$, the energy scale 
$\lambda^{1/4}\,m_P / \sqrt{\xi}$ during non-minimal SM inflation 
exceeds the effective ultraviolet cut-off scale  
$\Lambda = m_P/\xi$. 
Here $\lambda$ of order unity denotes the SM Higgs quartic coupling 
and $m_P \simeq 2.43 \times 10^{18}$~GeV represents 
the reduced Planck mass. 
Thus, the `flat' region of the effective potential lies 
beyond the region of applicability of the naive 
approximation, and so there is no compelling reason to trust 
the purported inflationary phase 
\cite{cutoff,Burgess:2010zq,Hertzberg:2010dc}. 
For a different viewpoint see Ref. \cite{Lerner:2009na}.

In this paper we reconsider non-minimal $\lambda\,\phi^4$ inflation 
and begin by replacing the SM Higgs inflaton with a gauge singlet 
scalar field. 
[The radial component of the axion field provides a nice example 
of a gauge singlet field and axion physics also 
provides a viable dark matter candidate.] 
We impose from the outset the requirement that the energy scale 
of inflation should not exceed the effective cut-off scale $\Lambda$. 
We also take into account quantum corrections to the inflationary 
potential arising from the interactions of the inflaton 
with other fields. 
Since one of our main goals is to obtain 
a lower bound on $r$, we only include corrections arising from 
the Yukawa interactions which can decrease $r$. 
We find that $r \gtrsim 0.002$, provided the scalar spectral 
index $n_s \geq 0.96$.
More generally, in this non-minimal $\lambda \phi^4$ inflation model, 
 the predictions for $n_s$ and $r$ lie within the WMAP 1-$\sigma$ 
 bounds for $10^{-12} \lesssim \lambda \lesssim 10^{-4}$ and 
 $10^{-3} \lesssim \xi \lesssim 10^2$. 
Recall that the corresponding tree level predictions for minimal 
($\xi = 0$) $\lambda\,\phi^4$ chaotic inflation, namely $n_s \simeq 0.95$ 
and $r \simeq 0.26$, lie outside the WMAP 2-$\sigma$ bounds.

We begin with the following tree level action in the Jordan frame:
\beq  \label{action1}
S_J^{tree} = \int d^4 x \sqrt{-g} 
\left[- \left( \frac{m_P^2 + \xi \phi^2}{2}\right)\mathcal{R}
+\frac{1}{2} (\partial \phi)^2 - \frac{\lambda }{4!} \phi^4 
-\frac{1}{2}y_N \phi \overline{N^c}N \right],
\eeq
where $\phi$ is a gauge singlet scalar field and $\lambda$ is 
the scalar self coupling. 
In order to keep the discussion simple, we have introduced 
only a single right handed neutrino $N$ with 
Yukawa coupling $y_N$, and we ignore the bare mass term for $N$. 
In a more realistic scenario,
at least two right-handed neutrinos are required 
for successful leptogenesis and reproducing 
neutrino oscillation data.

Using standard techniques \cite{RGEIP}, the one-loop 
renormalization group improved effective action can be written as
\beq
S_J = \int d^4 x \sqrt{-g} 
 \left[- \left( \frac{m_P^2 + \xi \phi^2}{2}\right)\mathcal{R}
+\frac{1}{2} (\partial \phi)^2 - \frac{1}{4!} \lambda(t) G(t)^4 \phi^4 \right],
\eeq
where $t=\ln(\phi/\mu)$ and $G(t)= \exp(- \, \int_0^t dt'
\gamma(t')/(1+\gamma(t')))$, with $\gamma(t) = \frac{y_N^2}{(4\pi)^2}$ 
being the anomalous dimension of the inflaton field. 
We ignore quantum corrections to the classical kinetic 
 and gravity sectors in the above action \cite{Barvinsky:2008ia,SHW}.
Moreover, as the inflaton is a gauge singlet field in our case, 
we only need to consider the RGEs of $\lambda$ and $y_N$:
\bea
\frac{d\lambda}{dt} &=& \frac{1}{(4\pi)^2} \left( 3 \lambda^2 
+ 4 \lambda \,y_N^2 -  24 y_N^4  \right),  \\
\frac{dy_N}{dt} &=& \frac{1}{(4\pi)^2} \left( \frac{5}{4} \, y_N^3  \right).
\eea
The requirement that the energy scale of inflation 
should lie below the cut-off scale ( $\Lambda = m_P/\xi$ for $\xi \geq 1$ and
$\Lambda = m_P$ for $\xi \leq 1 $) generates
values of the above couplings small enough to suppress the running of $\xi$.
Therefore, we ignore the running of $\xi$ in our numerical calculations.

In the Einstein frame with a canonical gravity sector, 
 the kinetic energy can be made canonical with respect 
 to a new field $\phie$ \cite{SHW}, 
\beq
\left({d\phie\over d\phi}\right)^{-2} =
 \frac{\left( 1 + \frac{\xi \phi^2}{m_P^2} \right)^2}{1+(6 \xi +1)\frac{\xi
 \phi^2}{m_P^2}}.  \label{kinetic}
\eeq
The action in the Einstein frame is then given by
\beq
S_E = \!\int d^4 x \sqrt{-g_E}\left[-\frac12 m_P^2 \mathcal{R}_E+\frac12 (\partial_E \phie)^2
-V_E(\phie(\phi))\right],
\label{action}
\eeq
with
\beq
V_E(\phi) = \frac{\frac{1}{4!} \lambda(t)\,G(t)^4\,\phi^4}{\left(1+\frac{\xi\,\phi^2}{m_P^2}\right)^2}.
\label{potrgi}
\eeq
To discuss things qualitatively it is convenient to use the following
approximate form of the above potential 
\beq
V_E(\phi) = \frac{\frac{1}{4!}\,\lambda\,\phi^4 
- \kappa\,\phi^4 \ln \left( \phi / \mu \right)}
{\left(1+\frac{\xi\,\phi^2}{m_P^2}\right)^2}, 
 \label{ApproxPotential}
\eeq
where we have assumed $\gamma \approx 0$, $dy_N/dt \approx 0$, $\lambda \ll y_N^2$, 
and $d \lambda /dt \approx - 24\,\kappa$ with $\kappa = y_N^4 /(4\pi)^2$. We have
checked that in the relevant parametric region the above potential can be considered
as a valid approximation. In our numerical calculations we fix the 
renormalization scale $\mu$ equal to the cut-off scale $\Lambda$.

Before starting our discussion of this model it is useful to 
recall here the basic results of the slow roll assumption. 
The inflationary slow-roll parameters are given by
\bea
\epsilon(\phi)&=&\frac12 m_P^2 \left({V_E'\over V_E \sigma'}\right)^2 
,\label{epsilon}\\
\eta(\phi)&=&m_P^2\left[
{V_E''\over V_E (\sigma')^2}
\!-{V_E'\sigma''\over V_E (\sigma')^3}\right],\,\,\,\,\,\label{eta}  \\
\zeta^2 (\phi) &=& m_P^4 \left({V_E'\over V_E \sigma'}\right) \left( \frac{V_E'''}{V_E (\sigma')^3}
-3 \frac{V_E'' \sigma''}{V_E (\sigma')^4} + 3 \frac{V_E' (\sigma'')^2}{V_E (\sigma')^5}
- \frac{V_E' \sigma'''}{V_E (\sigma')^4} \right)  , \label{xi}
\eea
where a prime denotes a derivative with respect to $\phi$. The slow-roll 
approximation is valid as long as the conditions 
$\epsilon \ll 1$, $|\eta| \ll 1$ and $\zeta^2 \ll 1$ hold. In this case 
the scalar spectral index $n_{s}$, the tensor-to-scalar ratio $r$, and the 
running of the spectral index $\frac{d n_{s}}{d \ln k}$ are 
approximately given by
\bea
n_s &\simeq& 1-6\,\epsilon + 2\,\eta, \\
r &\simeq& 16\,\epsilon,  \\ 
\frac{d n_{s}}{d \ln k} \!\!&\simeq&\!\!
16\,\epsilon\,\eta - 24\,\epsilon^2 - 2\,\zeta^2. \label{reqn}
\eea
The number of e-folds after the comoving scale $l$ has 
 crossed the horizon is given by
\beq
N_l={1\over \sqrt{2}\, m_P}\int_{\phi_{\rm e}}^{\phi_l}
{d\phi\over\sqrt{\epsilon(\phi)}}\left({d\phie\over d\phi}\right)\,,
\label{Ne}\eeq
where $\phi_l$ is the field value at the comoving scale 
$l$, and $\phi_e$ denotes the value of $\phi$ at
the end of inflation, defined by 
max$(\epsilon(\phi_e) , |\eta(\phi_e)|,\zeta^2(\phi_e)) = 1$.

The amplitude of the curvature perturbation $\Delta_{\mathcal{R}}$ is given by 
\begin{equation} \label{Delta}
\Delta_{\mathcal{R}}^2 = \left. \frac{V_E}{24\,\pi^2 \, m_P^2\,\epsilon } \right|_{k_0},
\end{equation}
where $\Delta_{\mathcal{R}}^2 = (2.43\pm 0.11)\times 10^{-9}$ is 
 the WMAP7 normalization at $k_0 = 0.002\, \rm{Mpc}^{-1}$ 
 \cite{Komatsu:2010fb}. 
Note that, for added precision, we include in our calculations 
 the first order corrections \cite{Stewart:1993bc} in the slow-roll expansion 
 for the quantities $n_s$, $r$, $\frac{dn_s}{d \ln k}$, 
 and $\Delta_{\mathcal{R}}$.

Using Eqs.~(\ref{ApproxPotential})-(\ref{Delta}) above 
we can obtain various predictions of the radiatively 
corrected non-minimal $\phi^4$ model of inflation. 
Once we fix the parameters $\xi$ and $\kappa$, 
and the number of e-foldings $N_0$, we can predict $n_s$, $r$, 
and $\frac{d n_{s}}{d \ln k}$. 
The tree level ($\kappa = 0$) minimal $\phi^4$ predictions 
are readily obtained as:
\bea
n_s &=& 1 - \frac{24}{\phi^2} = 1 - \frac{3}{N_0}, \\
r &=& \frac{128}{\phi^2} = \frac{16}{N_0}, \\
\frac{d n_{s}}{d \ln k}  &=& -\frac{192}{\phi^4} = -\frac{3}{N_0^2}.
\eea
For $N_0 = 60$, we find $n_s \simeq 0.95 $, $r \simeq 0.26$ 
and $\frac{d n_{s}}{d \ln k} \simeq -8 \times 10^{-3}$. 
As we mentioned above, this shows that the predictions of tree level minimal 
$\phi^4$ inflation lie outside the 2-$\sigma$ WMAP bounds \cite{Komatsu:2010fb}.
However, the situation improves once we include the radiative 
corrections \cite{NeferSenoguz:2008nn} generated from the Yukawa interaction 
in Eq. (\ref{action1}). Recently, these radiative corrections have been shown
to have important effects on the tree level predictions of various 
inflationary models \cite{Rehman:2009wv,Rehman:2010es}.
The scalar spectral index, the tensor to scalar ratio and the 
running of the spectral index for the radiatively corrected 
minimal $\phi^4$ inflation are then given by
\bea
n_s &\simeq& 1 - \left(  2\left( \frac{1-78\kappa /\lambda}{1-72\kappa /\lambda}\right)^2
-\left( \frac{1-86\kappa /\lambda}{1-72\kappa /\lambda}\right) \right) \frac{3}{N_0} , \\
r &\simeq& \left( \frac{1-78\kappa /\lambda}{1-72\kappa /\lambda}\right)^2\frac{16}{N_0}, \\
\frac{d n_{s}}{d \ln k}  &\simeq& -\frac{3}{N_0^2} 
\left( \frac{(1-98\kappa /\lambda)(1-78\kappa /\lambda)}{(1-72\kappa /\lambda)^2}  
\right) + \frac{r}{2} \left( \frac{16\,r}{3} - (1 - n_s ) \right) . 
\eea
\begin{figure}[t]
\centering \includegraphics[width=8.25cm]{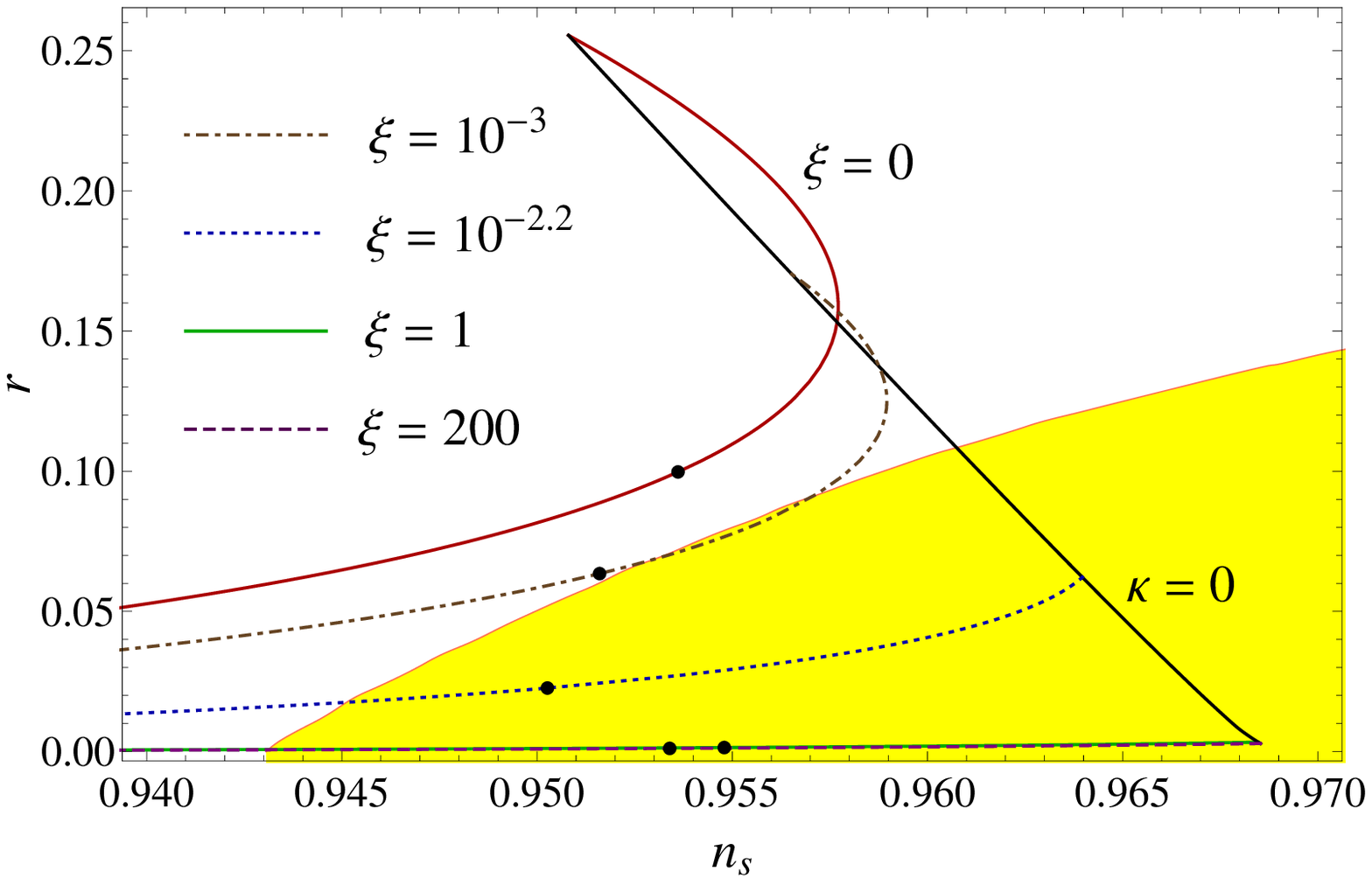}
\centering \includegraphics[width=8.25cm]{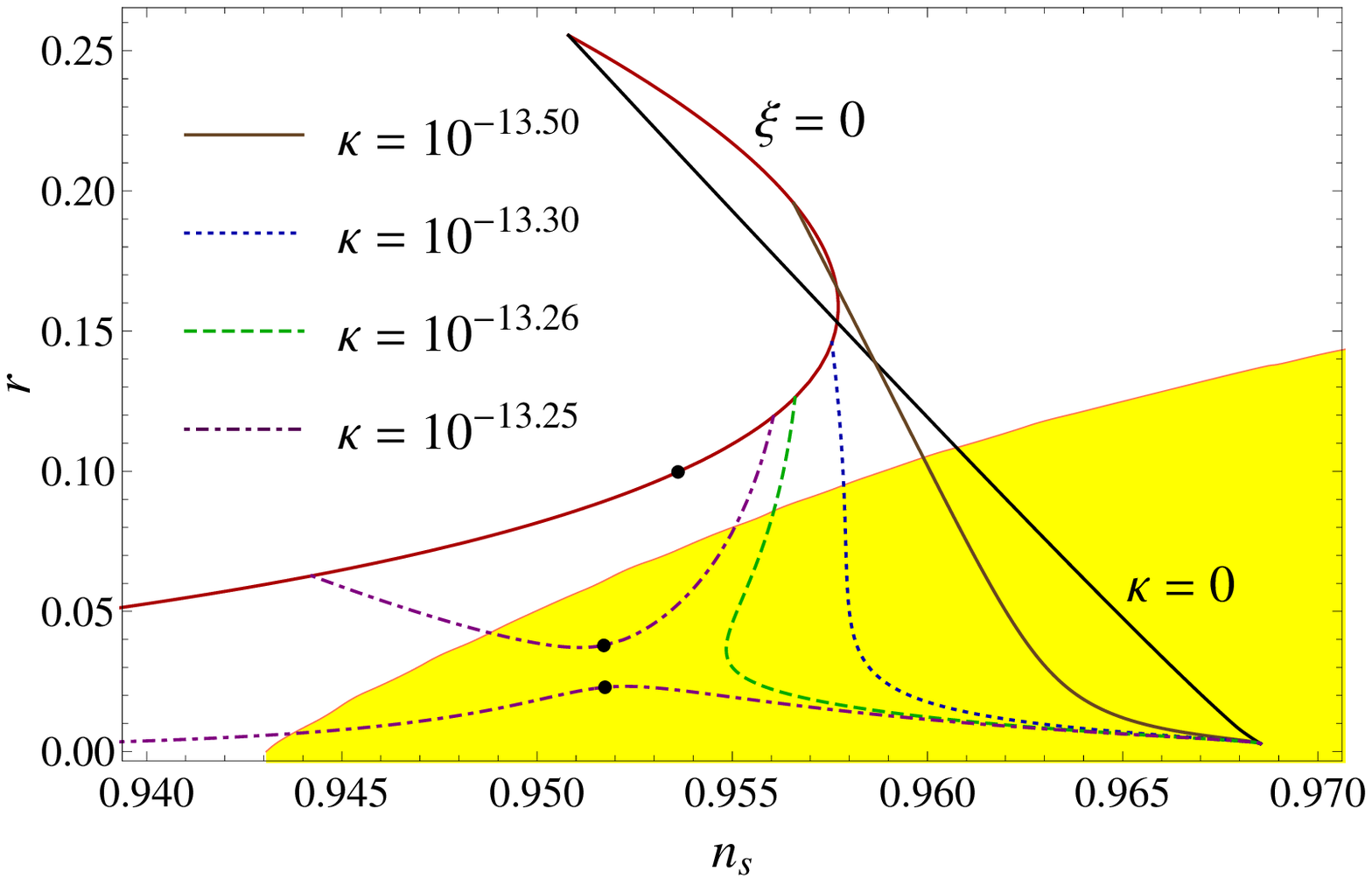}
\caption{$r$ vs. $n_s$ for the radiatively corrected 
non-minimal $\phi^4$ potential defined in Eq. (\ref{ApproxPotential})
with the number of e-foldings $N_0 = 60$. The WMAP 1-$\sigma$ 
(68\% confidence level) bounds are shown in yellow.
Along each curve we vary either $\kappa$ (left panel) or $\xi$ (right panel),
keeping one or the other fixed. The black dots represent the meeting 
points of the hilltop and the $\phi^4$ solutions
and correspond, for a given $\xi$, to the maximum value of $\kappa$.} \label{fig1}
\end{figure}

\begin{figure}[t]
\centering \includegraphics[width=8.25cm]{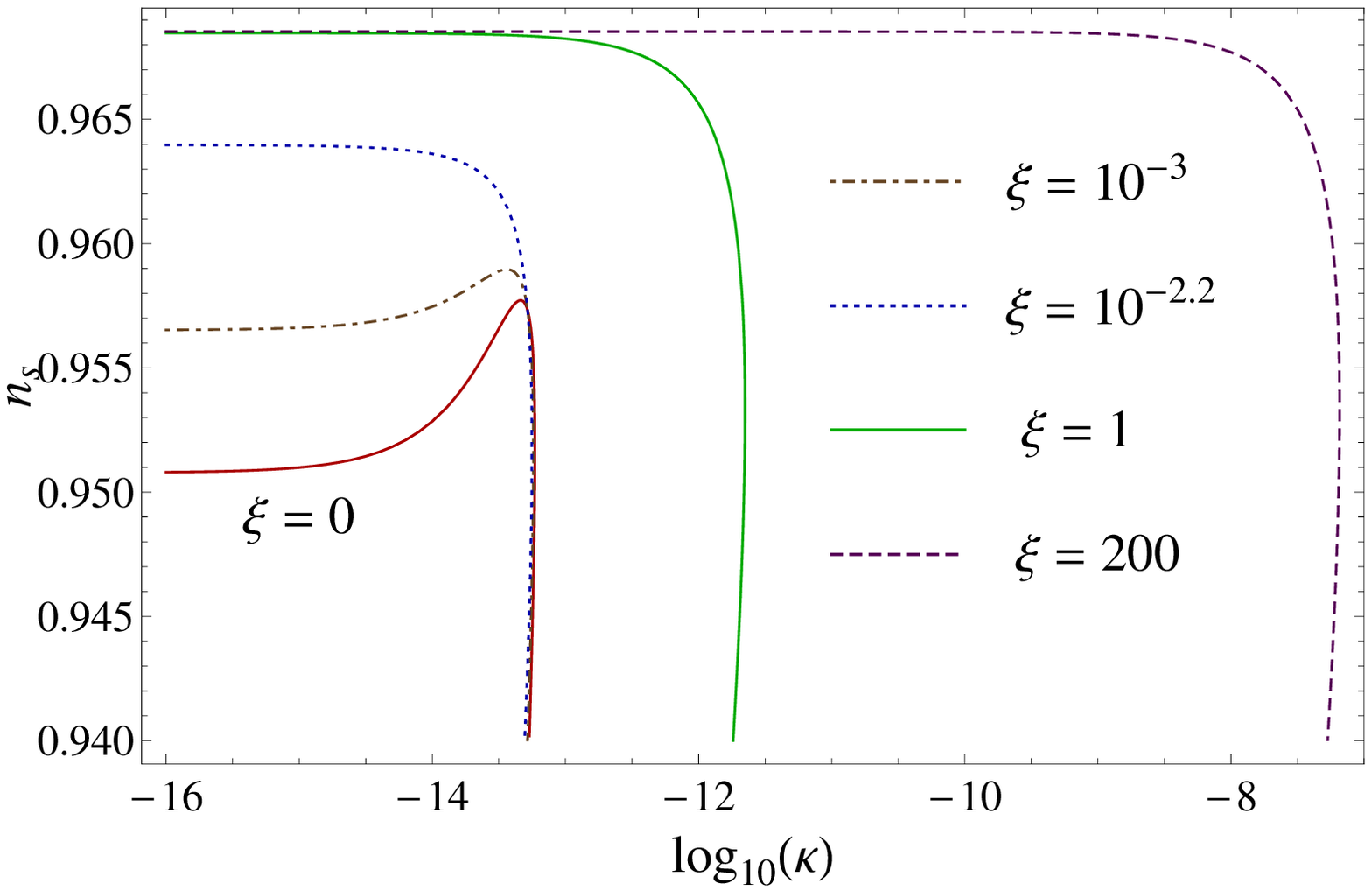}
\centering \includegraphics[width=8.25cm]{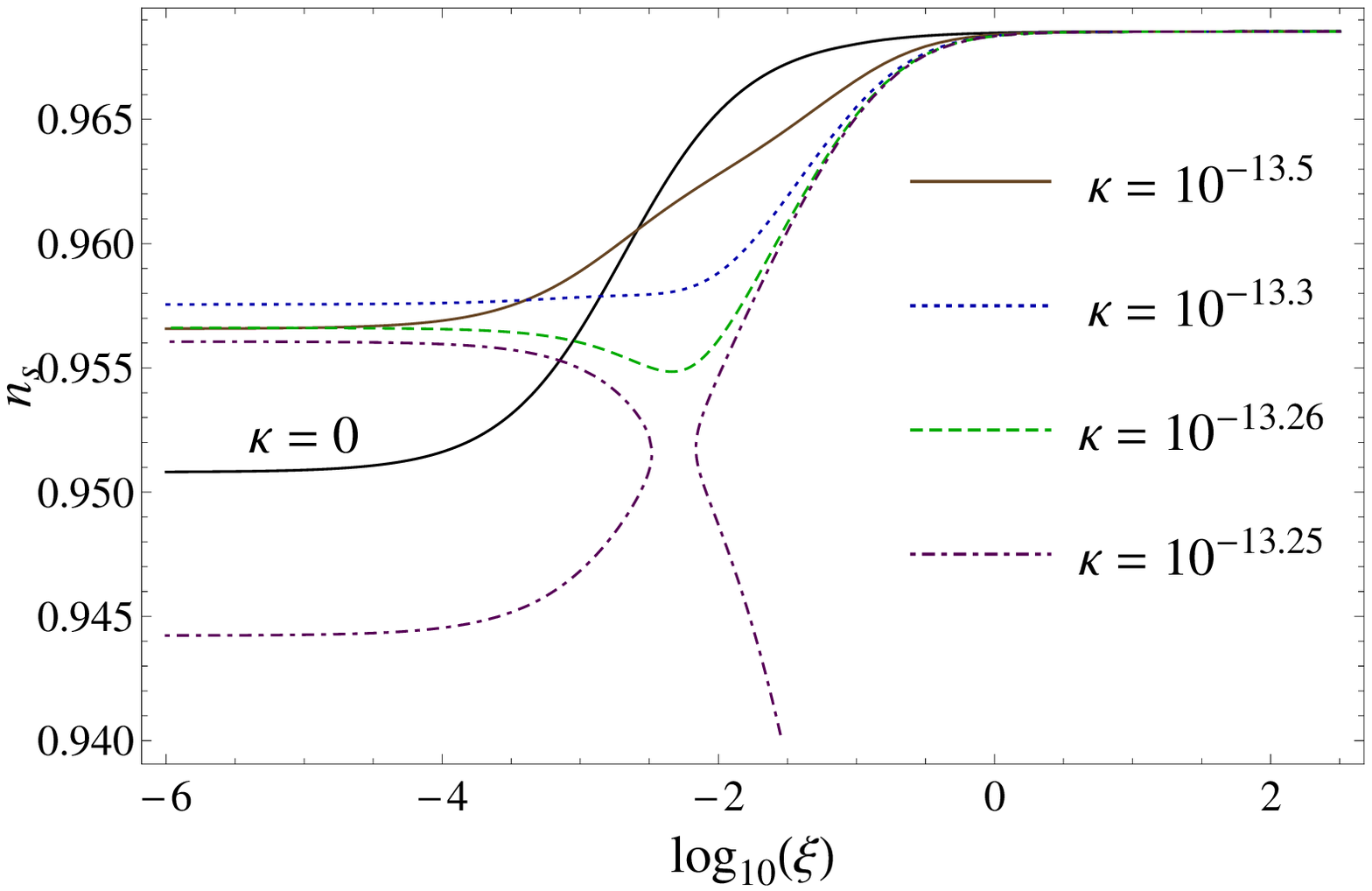}
\caption{$n_s$ vs. log$_{10}(\kappa)$ and log$_{10}(\xi)$ for  
radiatively corrected non-minimal $\phi^4$ inflation with 
the number of e-foldings $N_0 = 60$.} 
\label{fig2}
\end{figure}

\begin{figure}[h]
\centering \includegraphics[width=8.25cm]{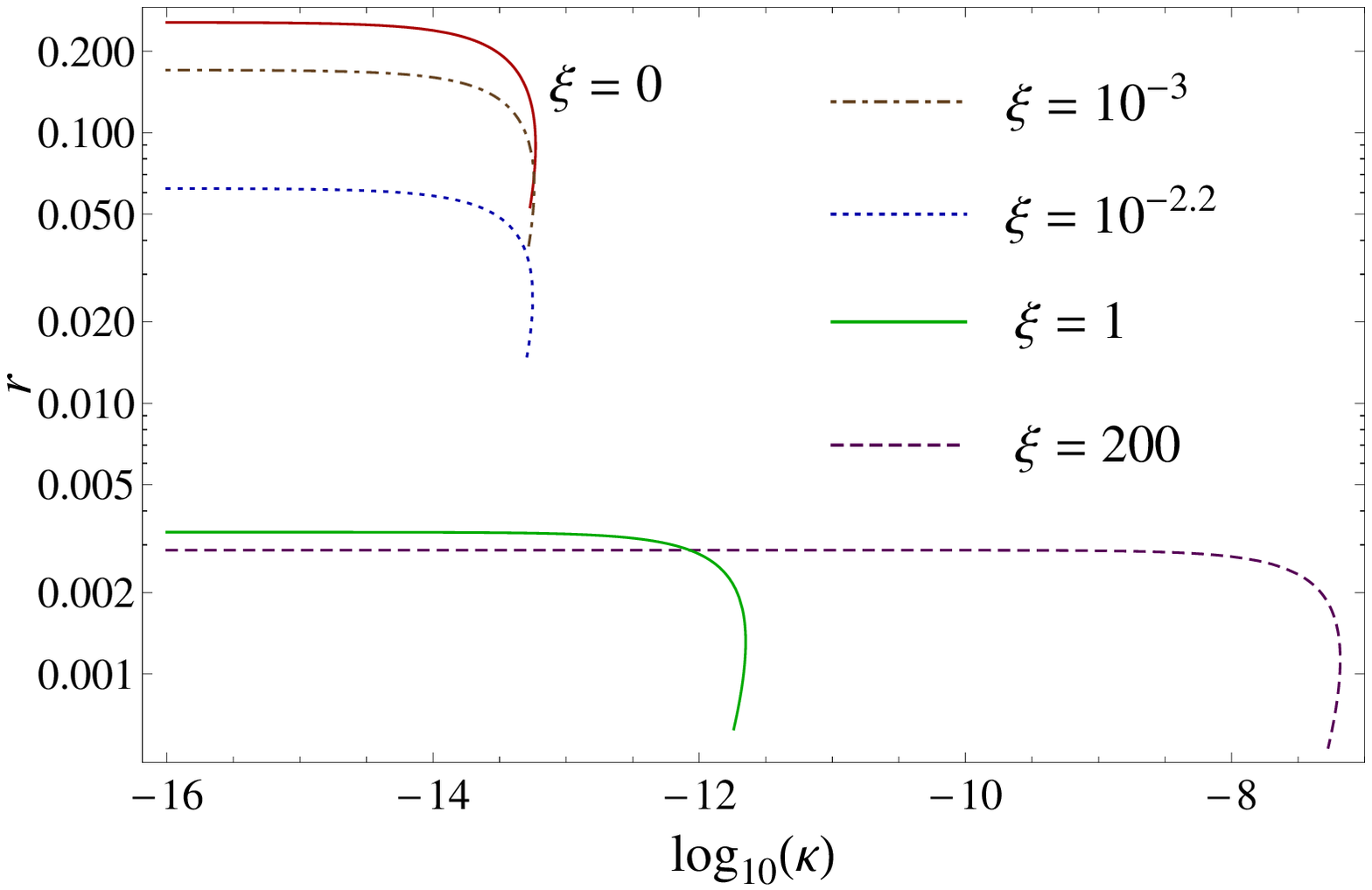}
\centering \includegraphics[width=8.25cm]{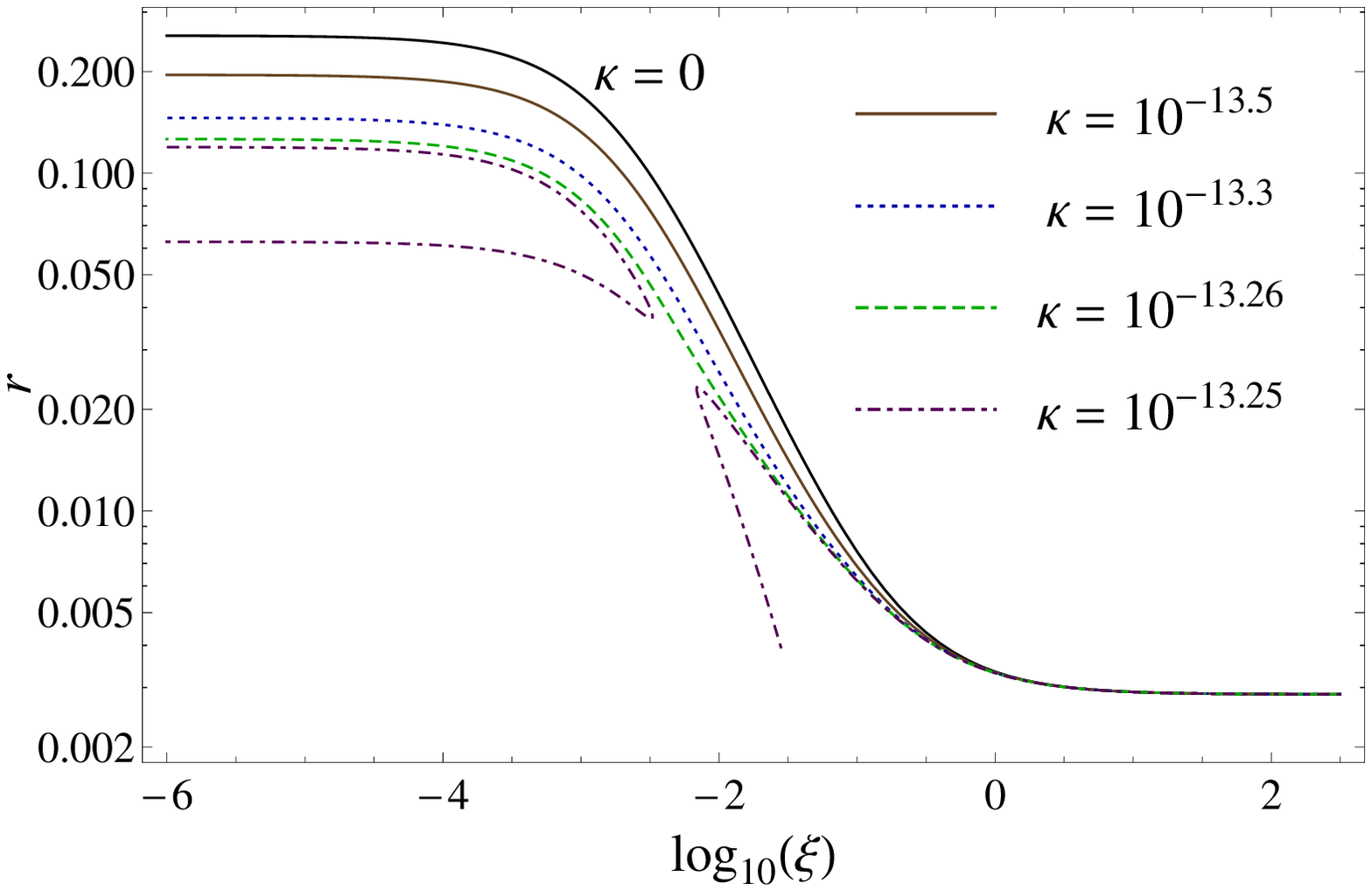}
\caption{$r$ vs. log$_{10}(\kappa)$ and log$_{10}(\xi)$ for 
radiatively corrected non-minimal $\phi^4$ inflation with 
the number of e-foldings $N_0 = 60$.} 
\label{fig3}
\end{figure}

The predicted values of $n_s$ and $r$ are shown in 
Figs.~\ref{fig1}-\ref{fig3} for $N_0 = 60$ e-foldings. 
The running of the spectral index $\frac{d n_{s}}{d \ln k}$ varies from 
$-3 \times 10^{-3}$ to $-8 \times 10^{-3}$. 
Although with the inclusion of radiative corrections 
we obtain a reduction in $r$, the predictions of the
radiatively corrected minimal $\phi^4$ inflation remain outside 
of the WMAP 1-$\sigma$ bounds. 
If we take $\kappa$ and the 
ratio $\kappa / \lambda$ as our two independent parameters 
(instead of $\kappa$ and $\lambda$), the value of $\kappa$
can be easily obtained in terms of $\kappa / \lambda$ and $N_0$
by employing Eq.~(\ref{Delta}):
\beq
\kappa \simeq \frac{(3\pi \Delta_{\mathcal{R}})^2}{N_0^3}
\frac{(\kappa / \lambda)(1-78 \kappa / \lambda)^2}{(1-72 \kappa / \lambda)^3}.
\eeq

\begin{figure}[t]
\centering \includegraphics[width=8.25cm]{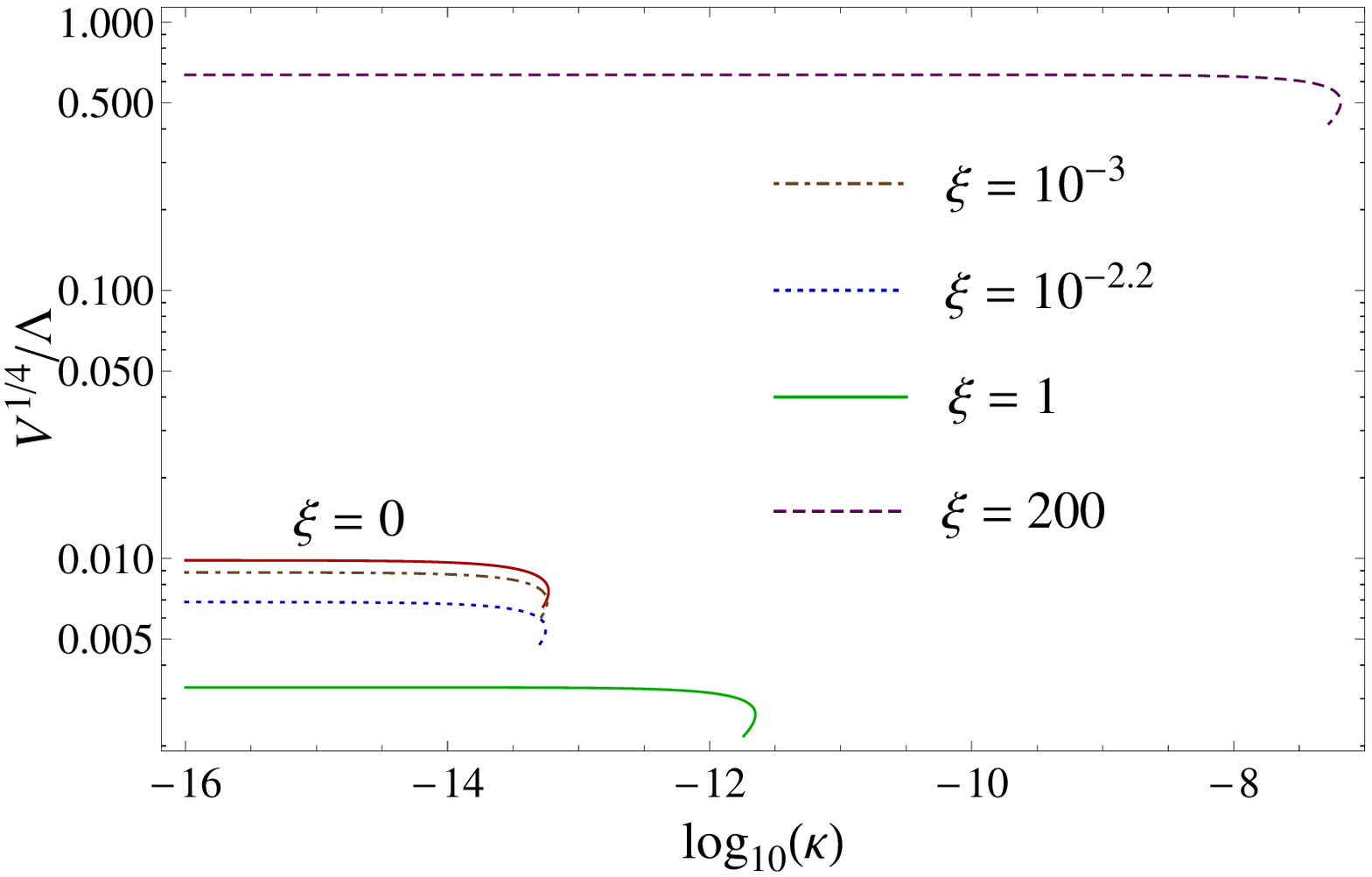}
\centering \includegraphics[width=8.25cm]{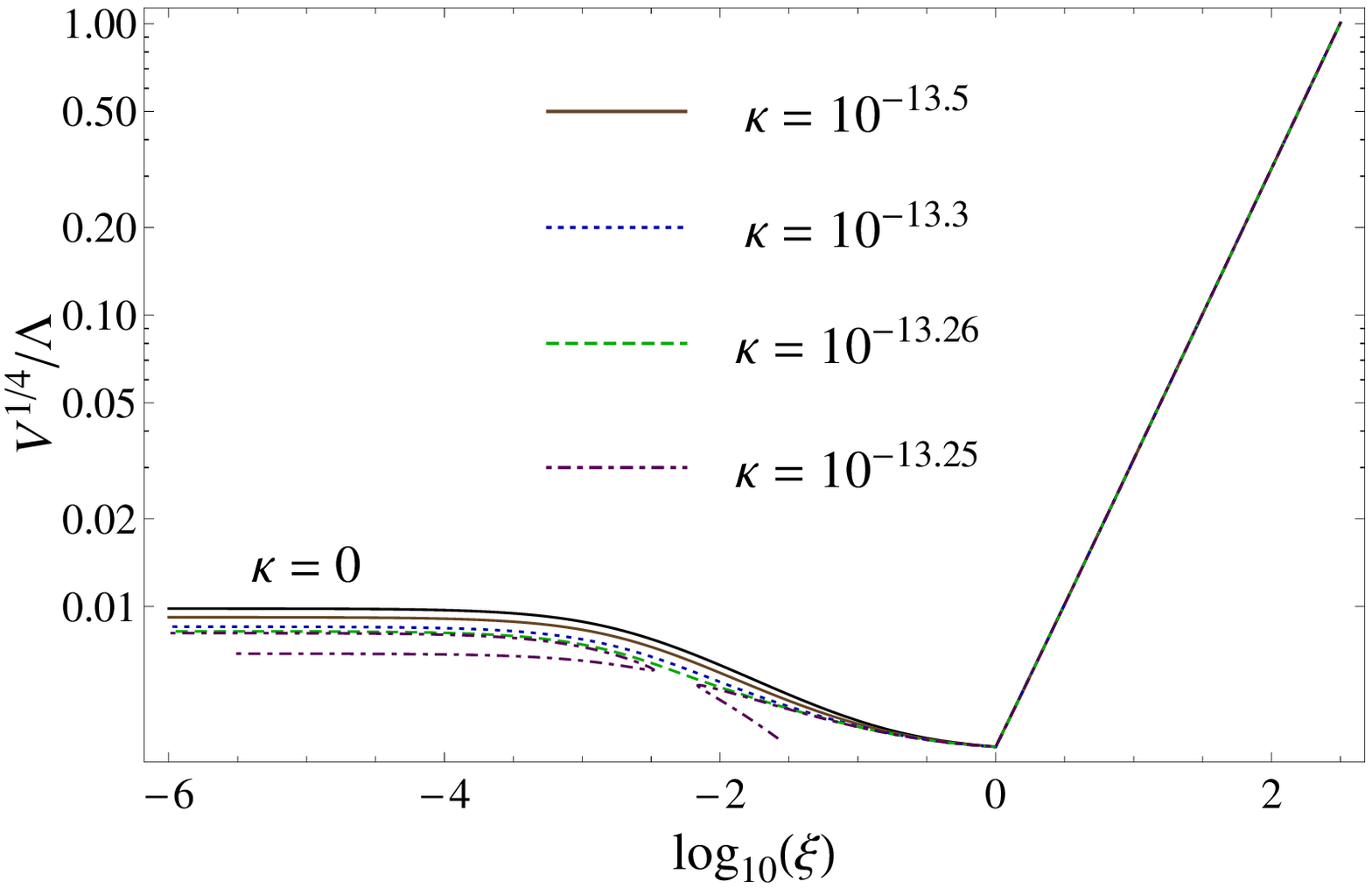}
\centering \includegraphics[width=8.25cm]{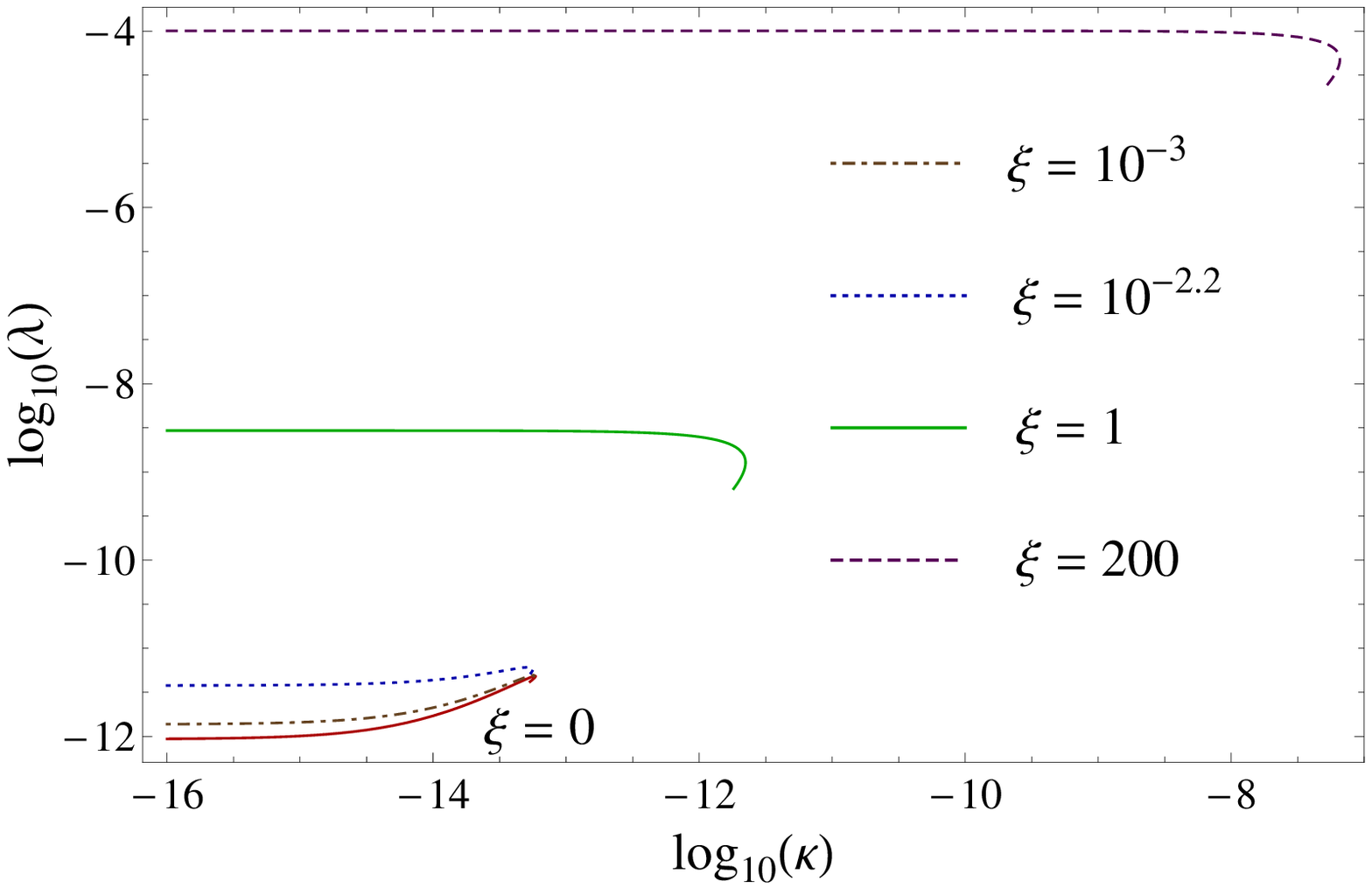}
\centering \includegraphics[width=8.25cm]{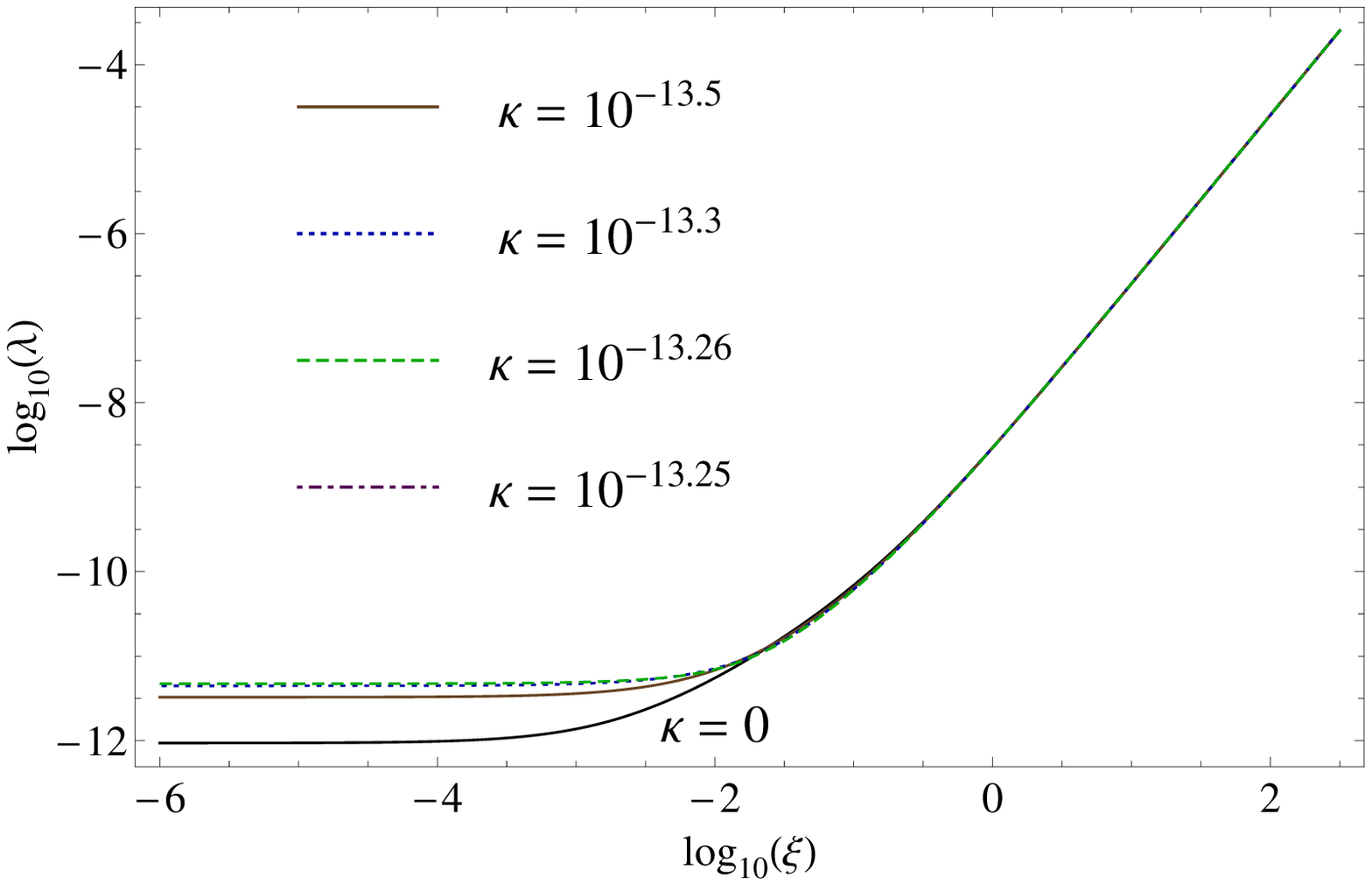}
\caption{$V^{1/4}/\Lambda$ and log$_{10}(\lambda)$ vs. 
log$_{10}(\xi)$ and log$_{10}(\kappa)$ for  
radiatively-corrected non-minimal $\phi^4$ inflation 
with the number of e-foldings $N_0 = 60$.} \label{fig4}
\end{figure}

The positive semi-definite condition $V \geq 0$ for the potential implies
$\kappa / \lambda  \leq 1/(24\ln(\phi/m_P)) \simeq 1/72$. However,
the WMAP 1-$\sigma$ bounds of the spectral index $n_s$ impose 
a more stringent bound $\kappa / \lambda \lesssim 1/79$. 
It is interesting to note that the above result allows two solutions 
for each of $\kappa / \lambda$, $n_s$ and $r$ for a given value of $\kappa$ 
\cite{NeferSenoguz:2008nn}. These two solutions meet at 
$\kappa/\lambda \sim 1/90$ and correspond to the maximum value of 
$\kappa \sim  10^{-13.6}$ as represented by the 
black dots in Fig.~\ref{fig1} and can be seen explicitly
in Figs.~\ref{fig3} and \ref{fig4}.
Following Ref.~\cite{NeferSenoguz:2008nn} 
we call the small $\kappa / \lambda$
solution the `$\phi^4$ solution' and 
the other the `hilltop solution'.
This hilltop solution mostly lies on the concave 
downward part of the potential, i.e. above the point of 
inflection whereas the $\phi^4$ solution lies below the point of 
inflection. Moreover, the value of the inflaton field at 
the pivot scale $\phi_0$ remains below the position of 
the hilltop in the WMAP 1-$\sigma$ region. 
In this letter we mainly restrict 
our discussion to the WMAP 1-$\sigma$ bounds.

For $\xi \neq 0$ and in the limit $\xi \ll 1$, the tree level predictions of minimal 
$\phi^4$ inflation are modified as follows:
\bea
n_s &\simeq&  1 - \frac{3(1+16\,\xi N_0/3)}{N_0\,(1+8\,\xi N_0)}, \\
r &\simeq&  \frac{16}{N_0\,(1+8\,\xi N_0)}, \\
\frac{d n_{s}}{d \ln k}  &\simeq& -\frac{3\,\left(1+4\,(8\,\xi N_0)/3-5\,(8\,\xi N_0)^2
-2\,(8\,\xi N_0)^3 \right)}{N_0^2\,(1+8\,\xi N_0)^4}
+ \frac{r}{2} \left(\frac{16\,r}{3} -(1 - n_s)\right).
\eea
These results exhibit a reduction in the value of $r$ and an increase
in the value of $n_s$ as can be seen in 
Figs.~\ref{fig1}-\ref{fig3}. In particular, from the WMAP 1-$\sigma$ 
bounds ($r \sim 0.1$ and $n_s \sim 0.96$), we obtain a lower bound of 
$\xi \gtrsim 3 \times 10^{-3}$ with $N_0 = 60$ e-foldings 
\cite{Bezrukov:2008dt}.
The tree level prediction for $\frac{d n_{s}}{d \ln k}$ 
receives only a tiny correction in this case. 
Note the sharp transitions in the predictions of $n_s$
and $r$ in the vicinity of $\xi \approx 10^{-2}$. This
can be understood from the expression for the inflationary
potential given in Eq.~(\ref{ApproxPotential}) and Eqs.~(24) and (25).

In order to discuss non-minimal $\phi^4$ inflation for 
$\xi \gg 1$, it is useful to define the dimensionless 
field variable $\psi \equiv \sqrt{\xi}\phi/m_P$. 
With $\xi$, $\psi \gg 1$, the tree level predictions
for $n_s$, $r$ and $\frac{d n_{s}}{d \ln k}$
are given by:
\bea
n_s &\simeq& 1 - \frac{8}{3\psi^2} = 1 - \frac{2}{N_0},  \\
r &\simeq& \frac{64}{3\psi^4} =  \frac{12}{N_0^2},  \\
\frac{d n_{s}}{d \ln k} &\simeq& -\frac{32}{9\psi^4} =  -\frac{2}{N_0^2},
\eea
with
\beq
\Delta_{\mathcal{R}}^2 \simeq \frac{\lambda}{\xi^2} \left( \frac{\psi^4}{768\,\pi^2} \right)
\simeq \frac{\lambda}{\xi^2}  \left( \frac{N_0^2}{432\,\pi^2}  \right). 
\eeq
The results are shown as a black curve in Figs.~\ref{fig1}-\ref{fig3}
labeled $\kappa = 0$.
The running of the spectral index 
$\frac{d n_{s}}{d \ln k} \simeq -5 \times 10^{-4}$ 
is somewhat smaller in comparison to the prediction 
of minimal $\phi^4$ inflation. 
The requirement that $V^{1/4} \lesssim \Lambda$ with $N_0 = 60$ 
e-foldings leads to the upper bounds $\xi \lesssim 300$ and 
$\lambda \lesssim 10^{-4}$ (see Fig.~\ref{fig4}).

The inclusion of radiative corrections modifies the tree level results 
of non-minimal $\phi^4$ inflation as follows:
\bea
n_s &\simeq& 1 - \frac{8}{3\psi^2} \left( \frac{1 + 6\kappa/\lambda (3 - 4\ln(\sqrt{\xi}\psi))}
{1 - 24\kappa/\lambda \ln(\sqrt{\xi}\psi)}\right), \\
r &\simeq&   \frac{64}{3\psi^4} \left( \frac{1 - 6\kappa/\lambda (\psi^2 +4\ln(\sqrt{\xi}\psi))}
{1 - 24\kappa/\lambda \ln(\sqrt{\xi}\psi)}\right)^2, \\
\frac{d n_{s}}{d \ln k} &\simeq& -\frac{32}{9\psi^4}
\left( \frac{(1 + 6\kappa/\lambda (5 - 4\ln(\sqrt{\xi}\psi)))
(1 - 6\kappa/\lambda (\psi^2 +4\ln(\sqrt{\xi}\psi)))}
{(1 - 24\kappa/\lambda \ln(\sqrt{\xi}\psi))^2}\right),
\eea
with
\beq
\Delta_{\mathcal{R}}^2 \simeq 
\frac{\lambda}{\xi^2} \left( \frac{\psi^4}{768\,\pi^2} \right)
\frac{\left(  1 - 24\kappa/\lambda \ln(\sqrt{\xi}\psi) \right)^3}
{\left(1 - 6\kappa/\lambda (\psi^2 +4\ln(\sqrt{\xi}\psi))\right)^2}.
\eeq
These results exhibit a reduction in the values of both $r$ and $n_s$ as
can be seen for the curves with $\xi = 200$ in Figs.~\ref{fig2} and \ref{fig3}.
In particular, for $n_s \geq 0.96 $ we obtain a lower bound 
$r \gtrsim 0.002$ (see Fig.~\ref{fig3}).
This may be compared with the result $r \gtrsim 0.02$ for the
Higgs potential found in Ref. \cite{Rehman:2010es}.
The running of the spectral index changes very slightly from 
$\frac{d n_{s}}{d \ln k} \sim -4 \times 10^{-4}$ to its 
tree level prediction $\frac{d n_{s}}{d \ln k} \sim -5 \times 10^{-4}$ 
within the WMAP 1-$\sigma$ bounds. For $\xi = 200$ the 
value of $\psi$ varies between $7$ and $9$.
The requirement that $V^{1/4} \lesssim \Lambda$ together 
with the WMAP 1-$\sigma$ bounds
implies an upper bound $\kappa \lesssim 10^{-7}$. 
The limiting case $\xi \ll 1$, on the other hand, shows 
similar trends for the scalar spectral index and the 
tensor to scalar ratio as can be seen, for example, with 
the $\xi = 10^{-3}$ curves in Figs.~\ref{fig2} and \ref{fig3}.

\begin{figure}[t]
\centering \includegraphics[width = 10cm]{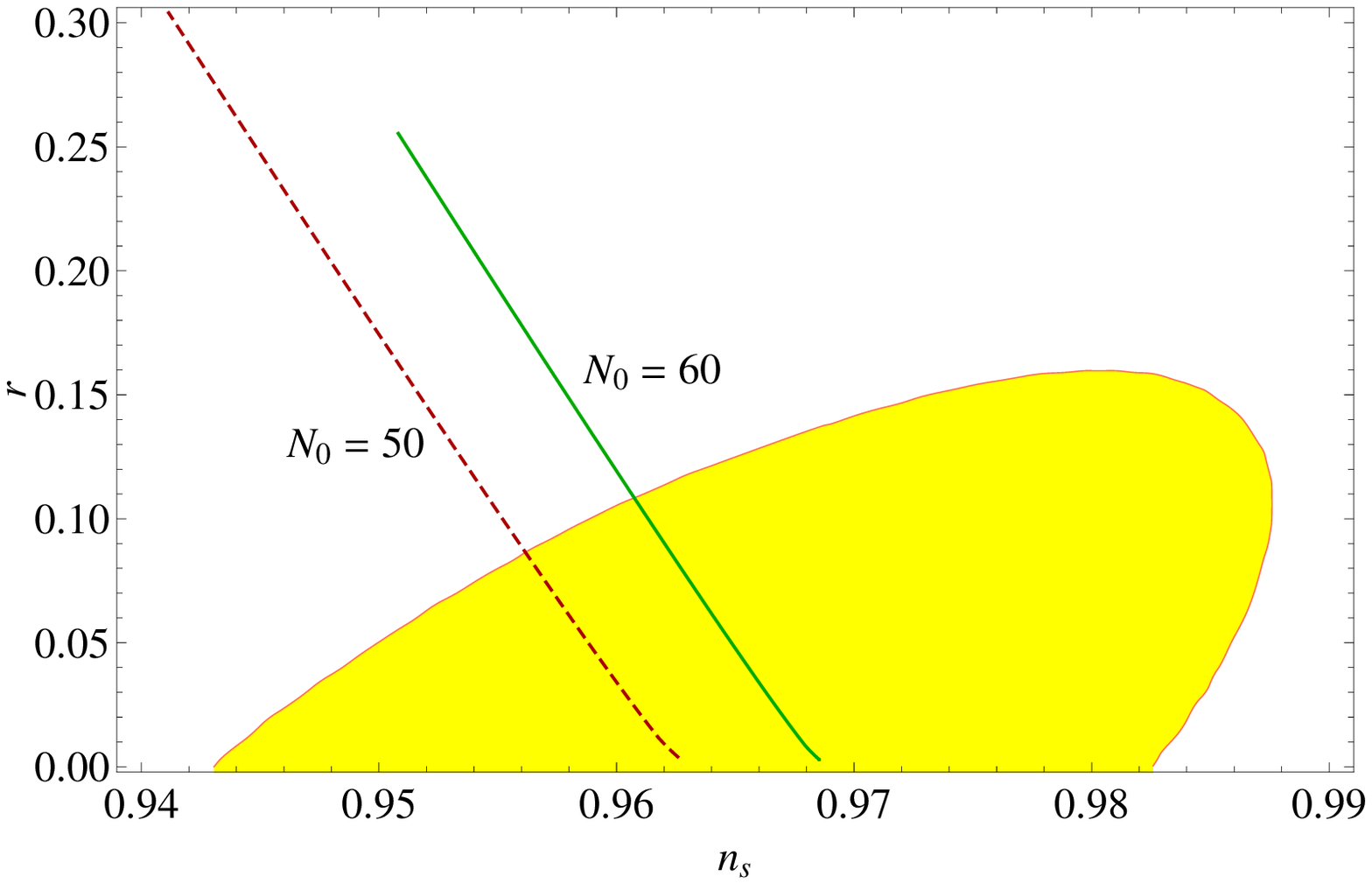}
\centering \includegraphics[width=8.25cm]{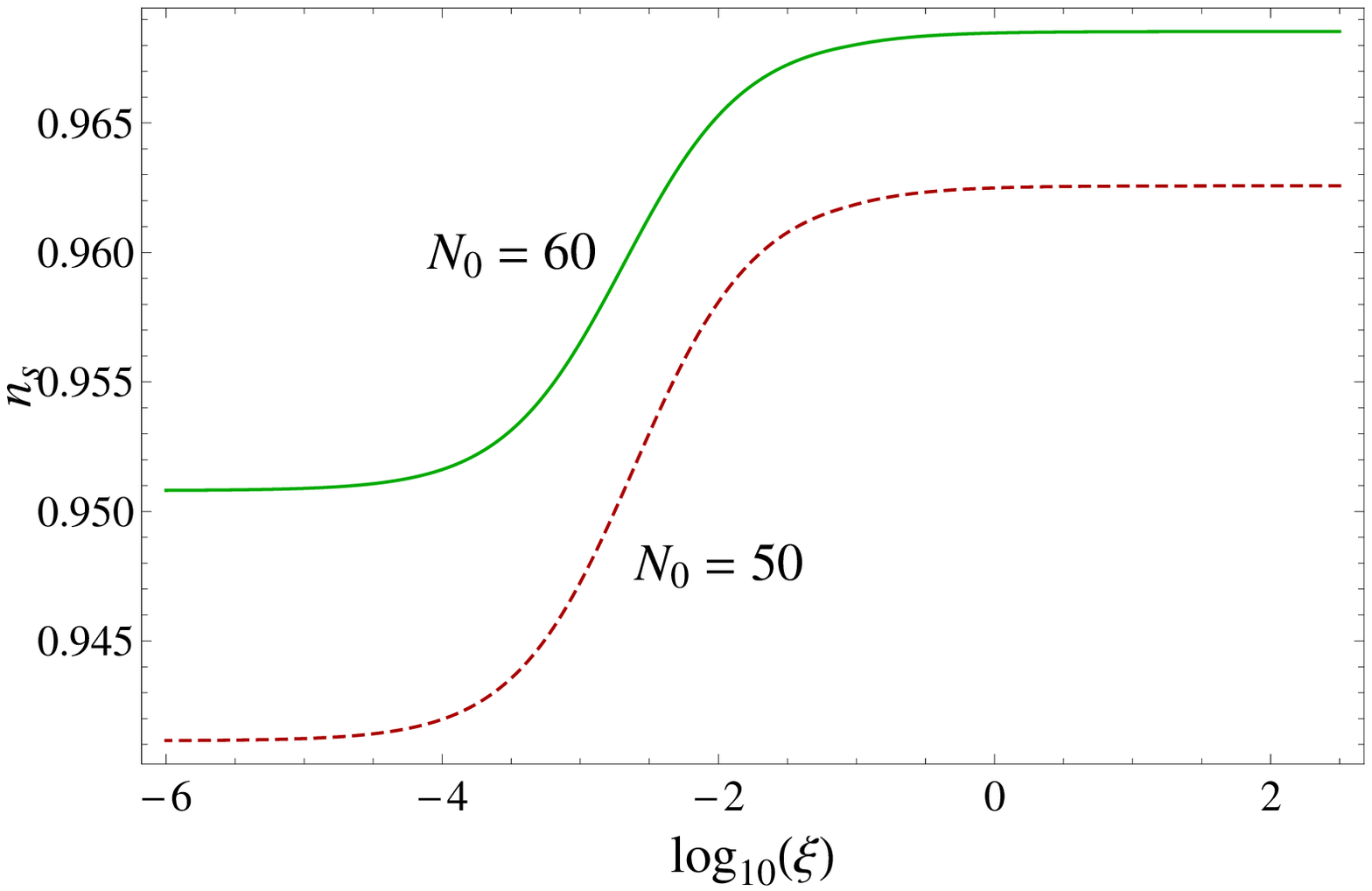}
\centering \includegraphics[width=8.25cm]{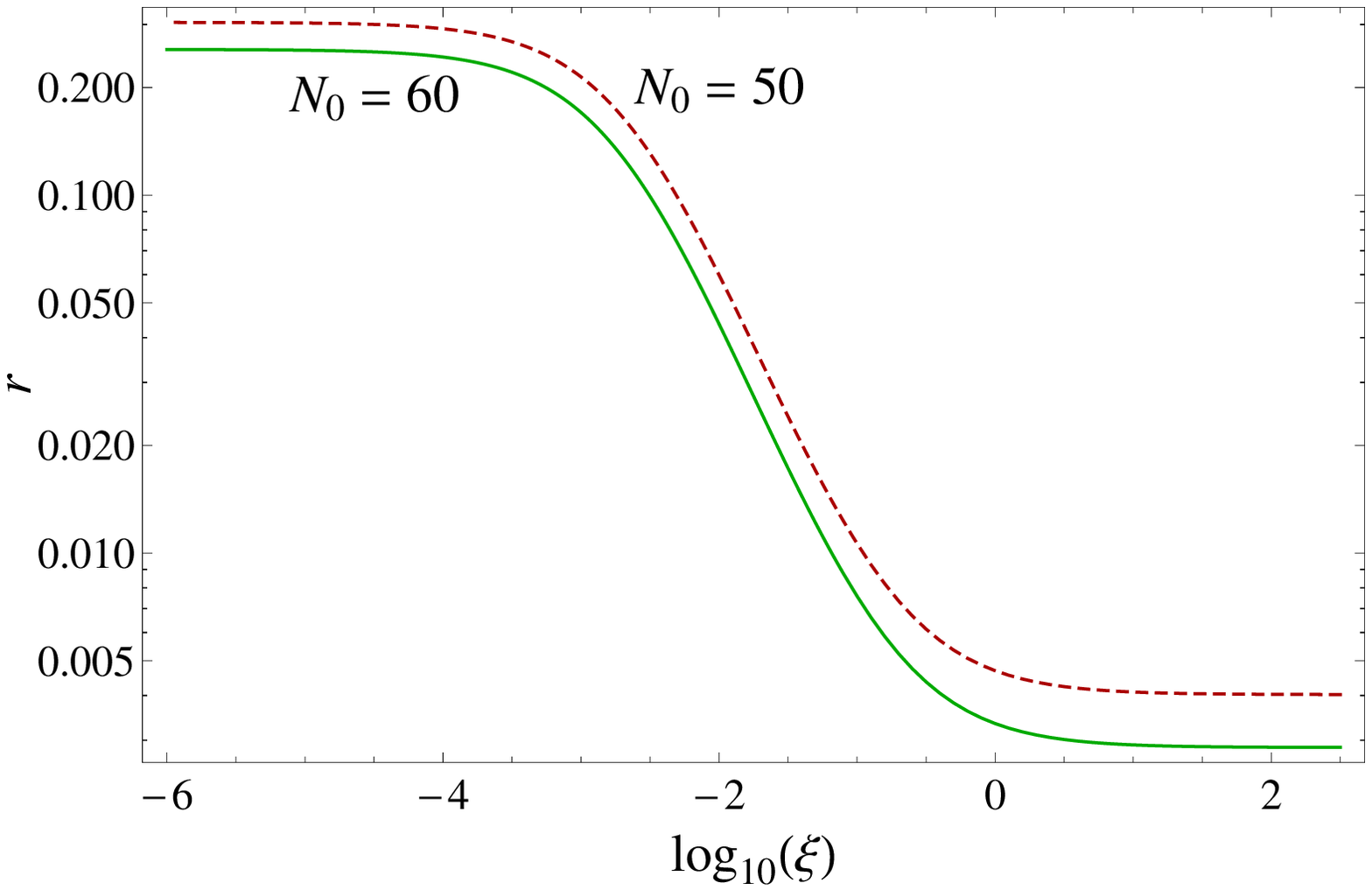}
\caption{$r$ vs. $n_s$ (first row) and $n_s$ and $r$ vs. log$_{10}(\xi)$ 
(second row) for tree level ($\kappa = 0$) non-minimal $\phi^4$ inflation 
with the number of e-foldings $N_0 = 50$ (red dashed curve) 
and $N_0 = 60$ (green solid curve). The WMAP 1-$\sigma$ 
(68\% confidence level) bounds are shown in yellow.} \label{fig5}
\end{figure}

\begin{figure}[t]
\centering \includegraphics[width=8.25cm]{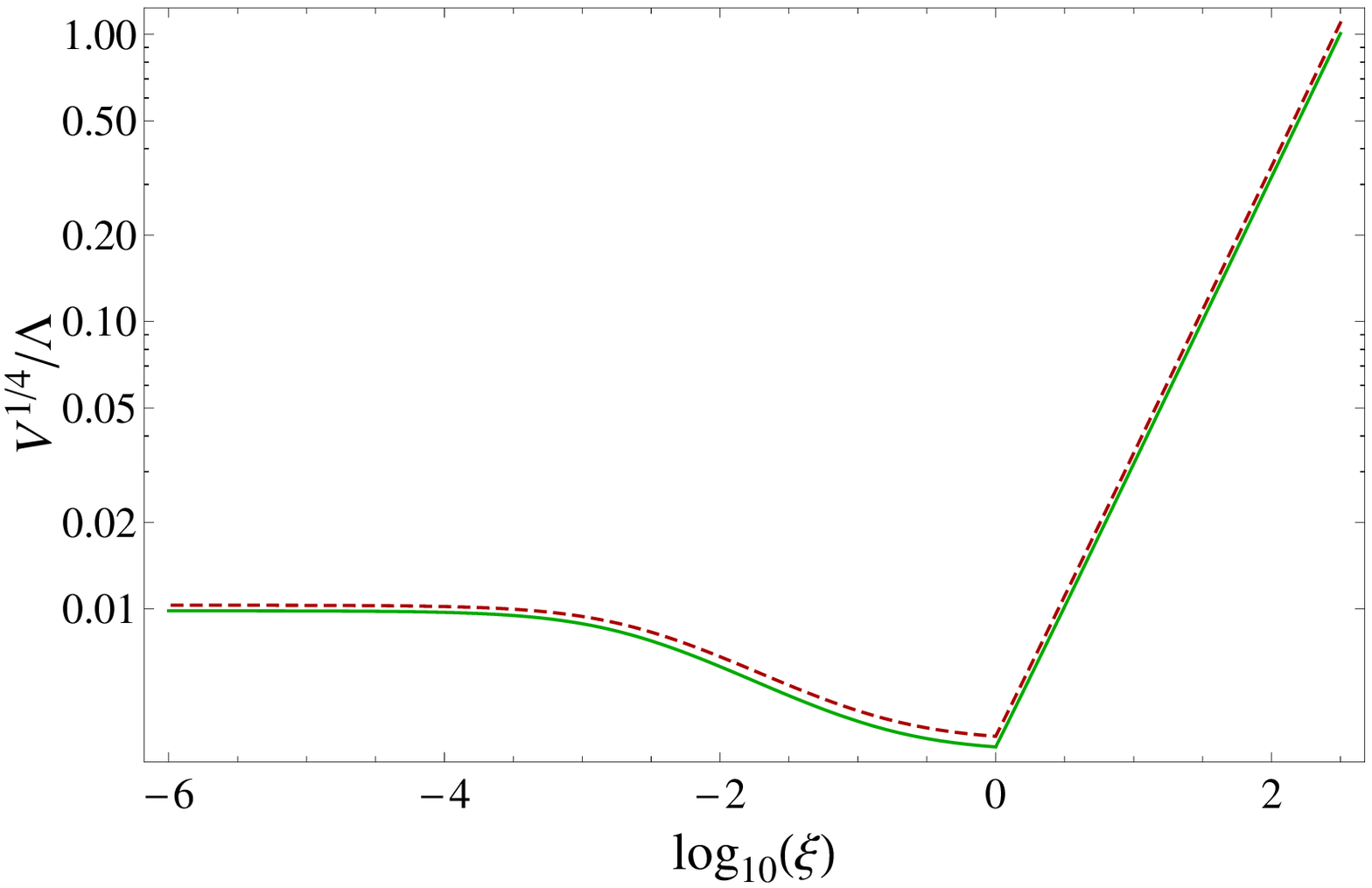}
\centering \includegraphics[width=8.25cm]{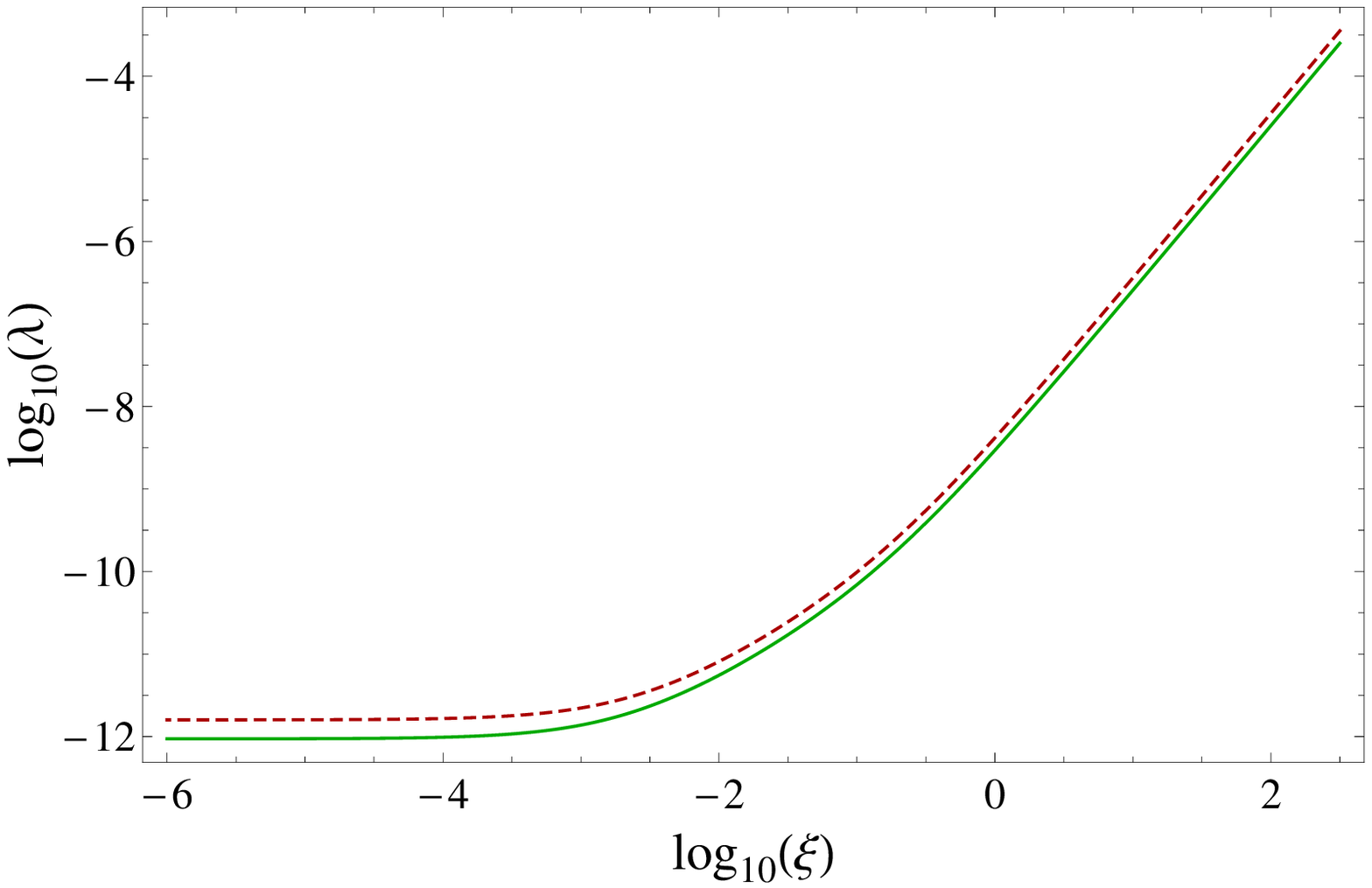}
\caption{$V^{1/4}/\Lambda$ and log$_{10}(\lambda)$ vs. log$_{10}(\xi)$ 
for tree level ($\kappa = 0$) non-minimal $\phi^4$ inflation 
with the number of e-foldings $N_0 = 50$ (red dashed curve) 
and $N_0 = 60$ (green solid curve).} \label{fig6}
\end{figure}

Finally in Figs.~\ref{fig5} and \ref{fig6} we display
the predictions of non-minimal $\phi^4$ inflation with
the number of e-foldings $N_0 = 50$ and $N_0 = 60$.
A reduction in $n_s$ and an increase in $r$ is observed
with a decrease in the number of e-foldings. This behavior 
is easy to understand with the help of analytical 
approximations derived in Eqs.~(27)-(28). The number of 
e-foldings $N_0 \simeq 50$ - $60$, depends on the 
reheating scenario. In our case, reheating occurs through 
the Yukawa coupling. Furthermore, the out of equilibrium 
decay of the inflaton can give rise to the observed 
baryon asymmetry via leptogenesis 
(either thermal~\cite{Fukugita:1986hr} or 
non-thermal~\cite{Lazarides:1991wu}).

To summarize, we have reconsidered non-minimal $\lambda\,\phi^4$ 
chaotic inflation and imposed the requirement that the energy 
scale of inflation remains below the effective UV cut-off 
scale i.e., $V^{1/4} \lesssim \Lambda$. 
The inflaton field $\phi$ is a gauge singlet scalar 
(say axion) field.
In addition to the non-minimal gravitational coupling, we have also 
included the Yukawa coupling of $\phi$ with a single right 
handed neutrino, leading to radiative corrections
which can have a significant effect. In the large $\xi \gg 1$ limit 
the requirement that $V^{1/4} \lesssim \Lambda$ provides the 
upper bounds $\xi \lesssim 10^2$, $\lambda \lesssim 10^{-4}$ and
$\kappa \lesssim 10^{-7}$, with predictions for $n_s$ and $r$
that are consistent with the WMAP 1-$\sigma$ bounds.
For $\xi \ll 1$, we obtain the lower bounds $\xi \gtrsim 10^{-3}$ 
and $\lambda \gtrsim 10^{-12}$ from the WMAP 1-$\sigma$ bounds.
Provided $n_s \geq 0.96$, we have shown that the scalar to
tensor ratio $r \gtrsim 0.002$, which will soon be 
tested by the Planck satellite.

\section*{Acknowledgments}
We thank Joshua R. Wickman for valuable discussions.
This work is supported in part by the DOE under grant 
No.~DE-FG02-91ER40626 (Q.S. and M.R.), and by the University of 
Delaware competitive fellowship (M.R.).



\end{document}